\documentclass{elsart}

\usepackage{graphicx}
\usepackage[dvips]{thumbpdf}
\usepackage{amssymb}
\usepackage{cite}

\begin{document}

\begin{frontmatter}

\title{Second-Order Eikonal Corrections for $A(e,e'p)$}

\author{B. Van Overmeire},
\author{J. Ryckebusch}\ead{Jan.Ryckebusch@UGent.be}

\address{Department of Subatomic and Radiation Physics,
Ghent University, Proeftuinstraat 86, B-9000 Gent, Belgium}

\begin{abstract}
The first-order eikonal approximation is frequently adopted in interpreting
the results of $A(e,e'p)$ measurements.  Glauber calculations, for example,
typically adopt the first-order eikonal approximation.  We present an
extension of the relativistic eikonal approach to $A(e,e'p)$ which accounts
for second-order eikonal corrections.  The numerical calculations are
performed within the relativistic optical model eikonal approximation.  The
nuclear transparency results indicate that the effect of the second-order
eikonal corrections is rather modest, even at $Q^{2} \approx 0.2$~(GeV/c)$^2$.
The same applies to polarization observables, left-right asymmetries, and
differential cross sections at low missing momenta.  At high missing momenta,
however, the second-order eikonal corrections are significant and bring the
calculations in closer agreement with the data and/or the exact results from
models adopting partial-wave expansions.
\end{abstract}

\end{frontmatter}

\section{Introduction}
\label{sec:intro}

The eikonal approximation \cite{mccauley58,glauber59,joachain75} has a long
history of successful results in describing scattering processes like
nucleon-nucleus scattering, heavy-ion collisions, and electroinduced
nucleon-knockout reactions.  The latter class of reactions, usually denoted
as $A(e,e'p)$, provide access to a wide range of nuclear phenomena like
short- and long-range correlations, relativistic effects, the transition
from hadronic to partonic degrees of freedom, and medium modifications of
nucleon properties.  The interpretation of $A(e,e'p)$ data heavily relies
on an accurate description of the effect of the final-state interactions
(\mbox{FSI}), i.e., the interactions of the ejected proton with the residual
nucleus such as rescattering and/or absorption.  The eikonal approximation
has been widely used to treat these distortions, either in combination with
optical potentials \cite{greenberg94,bianconi95,ito97,debruyne00}, or with
Glauber theory, its multiple-scattering extension
\cite{frankfurt95,nikolaev95,jeschonnek99,kohama00,benhar00,ciofi00,%
Petraki2003,ryckebusch03}.

The eikonal scattering wave functions are derived by linearizing the
continuum wave equation for the ejected proton.  Hence, the solution is
only valid to first order in $1/k$, with $k$ the proton's momentum, and
the eikonal approximation is suited for the description of reactions at
sufficiently high energies.  To extend the applicability to lower energies,
Wallace \cite{wallace} has developed systematic corrections to the eikonal
scattering amplitude.  Several authors have investigated the effect of
higher-order eikonal corrections in elastic nuclear scattering by protons,
antiprotons, and $\alpha$ particles \cite{waxman81,faldt92}, heavy-ion
collisions \cite{carstoiu93,cha95,alkhalili97,aguiar97}, and inclusive
electron-nucleus scattering \cite{tjon06}.  The aim of this Letter is to
determine the influence of higher-order eikonal corrections on $A(e,e'p)$
observables.  To this purpose, we extend the relativistic optical model
eikonal approximation (\mbox{ROMEA}) $A(e,e'p)$ framework of
Ref.~\cite{debruyne00}.  Our formalism builds upon the work of Baker
\cite{baker72}, where an eikonal approximation for potential scattering was
derived to second order in $1/k$.  Here, this work is extended to include the
effect of the spin-orbit potential.

The outline of this Letter is as follows.  In Section~\ref{sec:formalism},
the second-order eikonal correction to the \mbox{ROMEA} model is derived.
Section~\ref{sec:results} presents the results of the $A(e,e'p)$ numerical
calculations.  We look into how the second-order eikonal correction affects
more inclusive quantities like the nuclear transparency, as well as truly
exclusive observables such as the induced normal polarization $P_n$, the
left-right asymmetry $A_{LT}$, and the differential cross section.  Finally,
in Section~\ref{sec:concl}, we state our conclusions.

\section{Formalism}
\label{sec:formalism}

For the description of the $A(e,e'p)$ reaction, we adopt the impulse
approximation (IA) and the independent-nucleon picture.  Within this
approach, the basic quantity to be computed is the transition matrix
element \cite{kelly96}
\begin{equation}
\langle J^{\mu} \rangle =
\int d \vec{r} \,
\overline{\Psi}_{\vec{k},m_s}^{(-)} (\vec{r}) \,
\hat{J}^{\mu} (\vec{r}) \,
e^{i \vec{q} \cdot \vec{r}} \,
\phi_{\alpha_{1}} (\vec{r})
\; .
\label{eq:eep_matrix_elt}
\end{equation}
Here, $\phi_{\alpha_{1}}$ and $\Psi_{\vec{k},m_s}^{(-)}$ are the relativistic
bound-state and scattering wave functions, with $\alpha_{1}$ the quantum
numbers of the struck proton and $\vec{k}$ and $m_s$ the momentum and spin
of the ejected proton.  The relativistic bound-state wave function is
obtained in the Hartree approximation to the $\sigma - \omega$ model
\cite{serot86} with the W1 parametrization for the different field strengths
\cite{furnstahl97}.  The scattering wave function $\Psi_{\vec{k},m_s}^{(-)}$
appears with incoming boundary conditions and is related to
$\Psi_{\vec{k},m_s}^{(+)}$ by time reversal.  Furthermore, $\hat{J}^{\mu}$ is
the relativistic one-body current operator.  Throughout this Letter, we use
the Coulomb gauge and the CC2 form of $\hat{J}^{\mu}$ \cite{forest83}.

We now turn our attention to the determination of the scattering wave
function $\Psi_{\vec{k},m_s}^{(+)}$.  We start by considering the
Dirac equation for a proton with relativistic energy $E = \sqrt{k^{2}
+ M_N^{2}}$ and spin state $\left| \frac{1}{2} m_s \right>$ subject to
Lorentz scalar and vector potentials $V_{s} (r)$ and $V_{v} (r)$.  The Dirac
equation for the four-component spinor $\Psi_{\vec{k},m_s}^{(+)} (\vec{r})$
is converted to a Schr\"odinger-like equation for the upper component
$u^{(+)}_{\vec{k},m_s} (\vec{r})$ \cite{amado83,debruyne00}
\begin{equation}
\left[
-\frac{\nabla^2}{2M_N} + V_c(r) + V_{so}(r) \,
(\vec{\sigma} \cdot \vec{L} - i\vec{r} \cdot \hat{\vec{p}})
\right] \,
u_{\vec{k},m_s}^{(+)} (\vec{r}) =
\frac{k^2}{2M_N} \, u_{\vec{k},m_s}^{(+)} (\vec{r})
\; .
\label{eq:Schrodinger_upper}
\end{equation}
The central $V_{c} (r)$ and spin-orbit $V_{so} (r)$ potentials are defined
in terms of the scalar and vector ones, $V_{s} (r)$ and $V_{v} (r)$.  The
lower component $w^{(+)}_{\vec{k},m_s} (\vec{r})$ is related to the upper
one through
\begin{equation}
w_{\vec{k},m_s}^{(+)} (\vec{r}) =
\frac{1}{E+M_N+V_s(r)-V_v(r)} \,
\vec{\sigma} \cdot \hat{\vec{p}} \,
u_{\vec{k},m_s}^{(+)}(\vec{r})
\; .
\label{eq:lower_upper_relation}
\end{equation}
When solving Eq.~(\ref{eq:Schrodinger_upper}) in the eikonal approximation,
a standard procedure is to replace the momentum operator $\hat{\vec{p}}$ by
the asymptotic momentum $\vec{k}$ in the spin-orbit ($V_{so} (r) \,
\vec{\sigma} \cdot \vec{L}$) and Darwin ($V_{so} (r) \, (-i \vec{r} \cdot
\hat{\vec{p}})$) terms, as well as in the lower component
(\ref{eq:lower_upper_relation}).  In literature, this is usually referred
to as the effective momentum approximation (EMA) \cite{kelly99}.  For the
upper component, one puts forward a solution of the form
\begin{equation}
u_{\vec{k},m_s}^{(+)} (\vec{r}) \equiv
N \,
\eta (\vec{r}) \,
e^{i \vec{k} \cdot \vec{r}} \,
\chi_{\frac{1}{2}m_s}
\; ,
\label{eq:upper_ansatz}
\end{equation}
i.e., a plane wave modulated by an eikonal factor $\eta (\vec{r})$.  Here,
$N$ is a normalization factor.

In the \mbox{ROMEA} approach \cite{amado83,debruyne00}, which adopts the
first-order eikonal approximation, Eq.~(\ref{eq:Schrodinger_upper}) is
linearized in $\hat{\vec{p}}$ leading to a solution for the eikonal factor
of the form
\begin{equation}
\eta^{\mathrm{ROMEA}} (\vec{r}) =
\eta^{\mathrm{ROMEA}} (\vec{b}, z) =
\exp \left(
- i \, \frac{M_N}{k}
\int_{-\infty}^{z} dz \, ' \,
V_{\mathrm{opt}} (\vec{b}, z \, ')
\right)
\; ,
\label{eq:eikonal_phase}
\end{equation}
where $\vec{r} \equiv (\vec{b}, z)$, the $z$ axis lies along the momentum
$\vec{k}$ of the proton, and $V_{\mathrm{opt}} (\vec{b}, z) = V_{c}
(\vec{b}, z) + V_{so} (\vec{b}, z) \, (\vec{\sigma} \cdot \vec{b} \times
\vec{k} - i kz)$.  Despite the fact that it is written as an exponential
phase, the solution (\ref{eq:eikonal_phase}) is only valid up to first
order in $V_{\mathrm{opt}}/k$.

In what follows, we will derive an expression for the eikonal factor $\eta
(\vec{r})$ that is valid up to order $V_{\mathrm{opt}}/k^2$.  The momentum
dependence in the spin-orbit and Darwin terms makes that these terms are
retained up to order $V_{so}/k$, while central terms are included up to
order $V_{c}/k^2$.  Note that the expansion is not expressed in terms of
the Lorentz scalar and vector potentials $V_{s}$ and $V_{v}$.  Looking for
a solution of the form (\ref{eq:upper_ansatz}) for the Schr\"odinger-like
equation (\ref{eq:Schrodinger_upper}), Baker arrived at the following
equation for the eikonal factor (see Eq.~(14) of Ref.~\cite{baker72}):
\begin{eqnarray}
\eta (\vec{b}, z) & = &
1 -
i \, \frac{M_N}{k} \int_{-\infty}^{z} dz \, ' \,
V_{\mathrm{opt}} (\vec{b}, z \, ') \, \eta (\vec{b}, z \, ') +
\frac{M_N}{2k^2} \, V_{\mathrm{opt}} (\vec{b}, z) \, \eta (\vec{b}, z)
\nonumber \\ & & + \:
\frac{M_N}{2k^2} \int_{-\infty}^{z} dz \, ' \, (z - z \, ') \,
\left( \frac{1}{b} + \frac{\partial}{\partial b} \right)
\frac{\partial}{\partial b} \, 
\left( V_{\mathrm{opt}} (\vec{b}, z \, ') \,
\eta (\vec{b}, z \, ') \right)
\; .
\label{eq:2nd_order_eq_eik_factor}
\end{eqnarray}
Note that, apart from dropping contributions of order
$V_{\mathrm{opt}}/k^3$ and higher, no additional assumptions were made
when deriving Eq.~(\ref{eq:2nd_order_eq_eik_factor}).  In
Ref.~\cite{baker72}, Eq.~(\ref{eq:2nd_order_eq_eik_factor}) was
subsequently solved for spherically symmetric potentials.  The spin-orbit
and Darwin terms, however, break the spherical symmetry and a novel method
to solve Eq.~(\ref{eq:2nd_order_eq_eik_factor}) is needed.  To that purpose,
we assume that the derivative of the function $\eta$ is of higher order
in $1/k$ than $\eta$ itself (as is true for the \mbox{ROMEA} solution
(\ref{eq:eikonal_phase})).  This allows us to drop the $\partial \eta /
\partial b$ contribution in the last term of
Eq.~(\ref{eq:2nd_order_eq_eik_factor}), as it is of order
$V_{\mathrm{opt}}/k^3$ or higher:
\begin{eqnarray}
\lefteqn{
\frac{M_N}{2k^2} \int_{-\infty}^{z} dz \, ' \, (z - z \, ') \,
\left( \frac{1}{b} + \frac{\partial}{\partial b} \right)
\frac{\partial}{\partial b} \,
\left( V_{\mathrm{opt}} (\vec{b}, z \, ') \,
\eta (\vec{b}, z \, ') \right)
}
\nonumber \\ & = &
\frac{M_N}{2k^2}
\left( \frac{1}{b} + \frac{\partial}{\partial b} \right)
\int_{-\infty}^{z} dz \, ' \, (z - z \, ')
\nonumber \\ & & \times \:
\Biggl[
\frac{\partial}{\partial b} \, 
\left(
V_{c} (\vec{b}, z \, ') + V_{so} (\vec{b}, z \, ') \,
(\vec{\sigma} \cdot \vec{b} \times \vec{k} - i kz \, ')
\right)
\Biggr] 
\,
\eta (\vec{b}, z \, ')
\; .
\label{eq:last_term_eta_eq}
\end{eqnarray}
Spherical symmetry implies that $z \, ' \, \partial V_{c} (\vec{b}, z \,
') / \partial b = b \, \partial V_{c} (\vec{b}, z \, ') / \partial z
\, '$.  Hence, the $z \, ' \, \partial V_{c} / \partial b$ term in
Eq.~(\ref{eq:last_term_eta_eq}) can be written as
\begin{eqnarray}
\lefteqn{
- \frac{M_N}{2k^2}
\left( \frac{1}{b} + \frac{\partial}{\partial b} \right)
\int_{-\infty}^{z} dz \, ' \, b \,
\frac{\partial V_{c} (\vec{b}, z \, ')}{\partial z \, '} \,
\eta (\vec{b}, z \, ')
}
\nonumber \\ & = &
- \frac{M_N}{2k^2}
\left( \frac{1}{b} + \frac{\partial}{\partial b} \right)
\int_{-\infty}^{z} dz \, ' \, b \,
\frac{\partial}{\partial z \, '} \,
\left( V_{c} (\vec{b}, z \, ') \, \eta (\vec{b}, z \, ') \right)
\nonumber \\ & = &
- \frac{M_N}{2k^2} 
\Biggl[
\left( 2 + b \frac{\partial}{\partial b} \right) \, V_{c} (\vec{b}, z)
\Biggr]
\,
\eta (\vec{b}, z)
\; .
\label{eq:z_acc_central_term}
\end{eqnarray}
In the first step, we made use of the fact that the derivative $\partial
\eta / \partial z \, '$ is of higher order to turn the integrand into
an exact differential.  A similar reasoning, followed by integration by
parts, leads to
\begin{eqnarray}
\lefteqn{
\frac{M_N}{2k^2} 
\left( \frac{1}{b} + \frac{\partial}{\partial b} \right)
\int_{-\infty}^{z} dz \, ' \, (z - z \, ') \,
\frac{\partial V_{so} (\vec{b}, z \, ')}{\partial b} \,
(- i kz \, ') \,
\eta (\vec{b}, z \, ')
}
\nonumber \\ & = &
- i \, \frac{M_N}{2k}
\int_{-\infty}^{z} dz \, ' \,
\Biggl[
\left( 2 + b \frac{\partial}{\partial b} \right) \,
V_{so} (\vec{b}, z)
\Biggr]
\,
\eta (\vec{b}, z \, ')
\; ,
\label{eq:z_min_z_acc_Darwin_term}
\end{eqnarray}
for the Darwin term of Eq.~(\ref{eq:last_term_eta_eq}).  Inserting the
expressions of Eqs.~(\ref{eq:z_acc_central_term}) and
(\ref{eq:z_min_z_acc_Darwin_term}), Eq.~(\ref{eq:2nd_order_eq_eik_factor})
adopts the form
\begin{eqnarray}
\lefteqn{
\eta (\vec{b}, z) =
}
\nonumber \\ & &
1 -
i \, \frac{M_N}{k} \int_{-\infty}^{z} dz \, ' \,
V_{\mathrm{opt}} (\vec{b}, z \, ') \,
\eta (\vec{b}, z \, ')
- \frac{M_N}{2k^2}
\Biggl[
\left( 1 + b \frac{\partial}{\partial b} \right) \, V_{c} (\vec{b}, z)
\Biggr]
\,
\eta (\vec{b}, z)
\nonumber \\ & & + \:
\frac{M_N z}{2k^2 b} \left( 1 + b \frac{\partial}{\partial b} \right)
\int_{-\infty}^{z} dz \, ' \,
\frac{\partial V_{c} (\vec{b}, z \, ')}{\partial b} \,
\eta (\vec{b}, z \, ')
\nonumber \\ & & + \:
\frac{M_N}{2k^2} \,
V_{so} (\vec{b}, z) \,
(\vec{\sigma} \cdot \vec{b} \times \vec{k} - i kz) \,
\eta (\vec{b}, z)
\nonumber \\ & & + \:
\frac{M_N}{2k^2 b}
\left( 1 + b \frac{\partial}{\partial b} \right)
\int_{-\infty}^{z} dz \, ' \, (z - z \, ') \,
\Biggl[
\frac{\partial}{\partial b} \,
\left(
V_{so} (\vec{b}, z \, ') \, \vec{\sigma} \cdot \vec{b} \times \vec{k}
\right)
\Biggr]
\,
\eta (\vec{b}, z \, ')
\nonumber \\ & & - \:
i \, \frac{M_N}{2k}
\int_{-\infty}^{z} dz \, ' \,
\Biggl[
\left( 2 + b \frac{\partial}{\partial b} \right) \,
V_{so} (\vec{b}, z)
\Biggr]
\,
\eta (\vec{b}, z \, ')
\; .
\label{eq:2nd_order_eq_eik_factor_bis}
\end{eqnarray}

We look for a solution of the form
\begin{eqnarray}
\eta (\vec{b}, z) & = &
f (\vec{b}, z) \,
\exp \left(
- i \, \frac{M_N}{k}
\int_{-\infty}^{z} dz \, ' \,
V_{\mathrm{opt}} (\vec{b}, z \, ') \,
f (\vec{b}, z \, ')
\right)
\nonumber \\ & = & 
f (\vec{b}, z) \, \exp \left( i \, S (\vec{b}, z) \right)
\; ,
\label{eq:2nd_order_eikonal_phase}
\end{eqnarray}
which should reduce to the \mbox{ROMEA} result of Eq.~(\ref{eq:eikonal_phase})
when terms of higher order than $V_{\mathrm{opt}}/k$ are neglected.
Accordingly, the function $f (\vec{b}, z)$ should be of the form $f = 1 +
O (V_{\mathrm{opt}}/k^2)$.  Substituting (\ref{eq:2nd_order_eikonal_phase})
into Eq.~(\ref{eq:2nd_order_eq_eik_factor_bis}) and multiplying by $e^{- i
\, S (\vec{b}, z)}$ yields
\begin{eqnarray}
\lefteqn{
f (\vec{b}, z) =
1
- \frac{M_N}{2k^2}
\Biggl[
\left( 1 + b \frac{\partial}{\partial b} \right) \,
V_{c} (\vec{b}, z)
\Biggr]
\,
f (\vec{b}, z)
}
\nonumber \\ & & + \:
\frac{M_N z}{2k^2 b} \left( 1 + b \frac{\partial}{\partial b} \right)
\int_{-\infty}^{z} dz \, ' \,
\frac{\partial V_{c} (\vec{b}, z \, ')}{\partial b} \,
f (\vec{b}, z \, ')
\nonumber \\ & & + \:
\frac{M_N}{2k^2} \,
V_{so} (\vec{b}, z) \,
(\vec{\sigma} \cdot \vec{b} \times \vec{k} - i kz) \,
f (\vec{b}, z)
\nonumber \\ & & + \:
\frac{M_N}{2k^2 b}
\left( 1 + b \frac{\partial}{\partial b} \right)
\int_{-\infty}^{z} dz \, ' \, (z - z \, ') \,
\Biggl[
\frac{\partial}{\partial b} \,
\left(
V_{so} (\vec{b}, z \, ') \, \vec{\sigma} \cdot \vec{b} \times \vec{k}
\right)
\Biggr]
\,
f (\vec{b}, z \, ')
\nonumber \\ & & - \:
i \, \frac{M_N}{2k}
\int_{-\infty}^{z} dz \, ' \,
\Biggl[
\left( 2 + b \frac{\partial}{\partial b} \right) \,
V_{so} (\vec{b}, z)
\Biggr]
\,
f (\vec{b}, z \, ')
\; .
\label{eq:2nd_order_eq_f_function}
\end{eqnarray}
In deriving this equation, we set $e^{i \, S (\vec{b}, z \, ')} \,
e^{- i \, S (\vec{b}, z)}$ equal to $1$, since higher-order terms are
neglected.  The difficulty in solving for $f (\vec{b}, z)$ is that
Eq.~(\ref{eq:2nd_order_eq_f_function}) is an integral equation.  An
expression for $f (\vec{b}, z)$ can, however, be readily obtained by
adding $(1 - f)$ terms, which introduce only higher-order terms, to the
right-hand side of Eq.~(\ref{eq:2nd_order_eq_f_function}).  This is
permitted since we seek for a solution up to order $V_{\mathrm{opt}}/k^2$.
With this manipulation, the function $f$ becomes
\begin{eqnarray}
\lefteqn{
f (\vec{b}, z) =
1
- \frac{M_N}{2k^2}
\left( 1 + b \frac{\partial}{\partial b} \right) \,
V_{c} (\vec{b}, z)
+ \frac{M_N z}{2k^2 b} \left( 1 + b \frac{\partial}{\partial b} \right)
\int_{-\infty}^{z} dz \, ' \,
\frac{\partial V_{c} (\vec{b}, z \, ')}{\partial b}
}
\nonumber \\ & & + \:
\frac{M_N}{2k^2} \,
V_{so} (\vec{b}, z) \,
(\vec{\sigma} \cdot \vec{b} \times \vec{k} - i kz)
\nonumber \\ & & + \:
\frac{M_N}{2k^2 b}
\left( 1 + b \frac{\partial}{\partial b} \right)
\int_{-\infty}^{z} dz \, ' \, (z - z \, ') \,
\frac{\partial}{\partial b} \,
\left(
V_{so} (\vec{b}, z \, ') \, \vec{\sigma} \cdot \vec{b} \times \vec{k}
\right)
\nonumber \\ & & - \:
i \, \frac{M_N}{2k}
\int_{-\infty}^{z} dz \, ' \,
\left( 2 + b \frac{\partial}{\partial b} \right) \,
V_{so} (\vec{b}, z)
\,
\; .
\label{eq:f_function}
\end{eqnarray}
The eikonal factor of Eq.~(\ref{eq:2nd_order_eikonal_phase}) with
$f (\vec{b}, z)$ given by (\ref{eq:f_function}), is a solution of the
integral equation (\ref{eq:2nd_order_eq_eik_factor}) to order
$V_{\mathrm{opt}}/k^2$ and reduces to the \mbox{ROMEA} result
(\ref{eq:eikonal_phase}) when truncated at order $V_{\mathrm{opt}}/k$.
Furthermore, it can be easily verified that the derivative of $\eta$ is
of higher order in $V_{\mathrm{opt}}/k$ than $\eta$ itself.  Henceforth,
calculations performed with the eikonal factor of
Eqs.~(\ref{eq:2nd_order_eikonal_phase}) and (\ref{eq:f_function}), are
dubbed as the second-order relativistic optical model eikonal
approximation (\mbox{SOROMEA}).

\section{Results}
\label{sec:results}

One way to quantify the overall effect of \mbox{FSI} in $A(e,e'p)$ processes
is via the nuclear transparency.  The measurements are commonly performed
under quasielastic conditions
\cite{garino92,oneill95,makins94,abbott98,dutta03,rohe05}.  We obtain the
theoretical transparencies by adopting similar expressions and cuts as
in the experiments.  Hence, the nuclear transparency is defined as
\cite{lava04}
\begin{equation}
T =
\frac{\sum_{\alpha} \int_{\Delta^3 p_m} d\vec{p}_m
S^{\alpha} (\vec{p}_m,E_m,\vec{k})}
{c_A \sum_{\alpha} \int_{\Delta^3 p_m} d\vec{p}_m
S^{\alpha}_{\mathrm{PWIA}} (\vec{p}_m,E_m)}
\; .
\label{eq:T_eep_theo}
\end{equation}
Here, $S^{\alpha}$ is the reduced cross section for knockout from the
shell $\alpha$
\begin{equation}
S^{\alpha}(\vec{p}_m,E_m,\vec{k}) =
\frac{
\frac{d^5 \sigma^{\alpha}}{d \Omega_p d \epsilon ' d \Omega_{\epsilon '}}
(e,e'p)}{K \sigma_{ep}} 
\; , 
\label{eq:eep_reduced_cross}
\end{equation}
where $\vec{p}_m$ and $E_m$ are the missing momentum and energy, $K$ is
a kinematical factor and $\sigma_{ep}$ is the off-shell electron-proton
cross section.  $S^{\alpha}_{\mathrm{PWIA}}$ is the reduced cross section
within the plane-wave impulse approximation (\mbox{PWIA}) in the
nonrelativistic limit.  Further, $\sum_{\alpha}$ extends over all occupied
shells $\alpha$ in the target nucleus.  The phase-space volume in the
missing momentum $\Delta^3 p_m$ is defined by the cut $|p_m| \leq 300$~MeV/c.
The $A$-dependent factor $c_A$ corrects in a phenomenological way for the
effect of short-range correlations.  We introduce the $c_A$ in the
denominator of Eq.~(\ref{eq:T_eep_theo}) because the data have undergone
a rescaling with $c_A = 0.9$ (\nuc{12}{C}) and $0.82$ (\nuc{56}{Fe}).

Transparencies have been computed for the nuclei \nuc{12}{C} and
\nuc{56}{Fe} at planar and constant $(\vec{q}, \omega)$ kinematics
compatible with the phase space covered in the experiments.  For the
optical potential, the \mbox{EDAD1} parametrization of
Ref.~\cite{cooper93} was used.

\begin{figure*}[p]
\begin{center}
\includegraphics[width=0.75\textwidth]{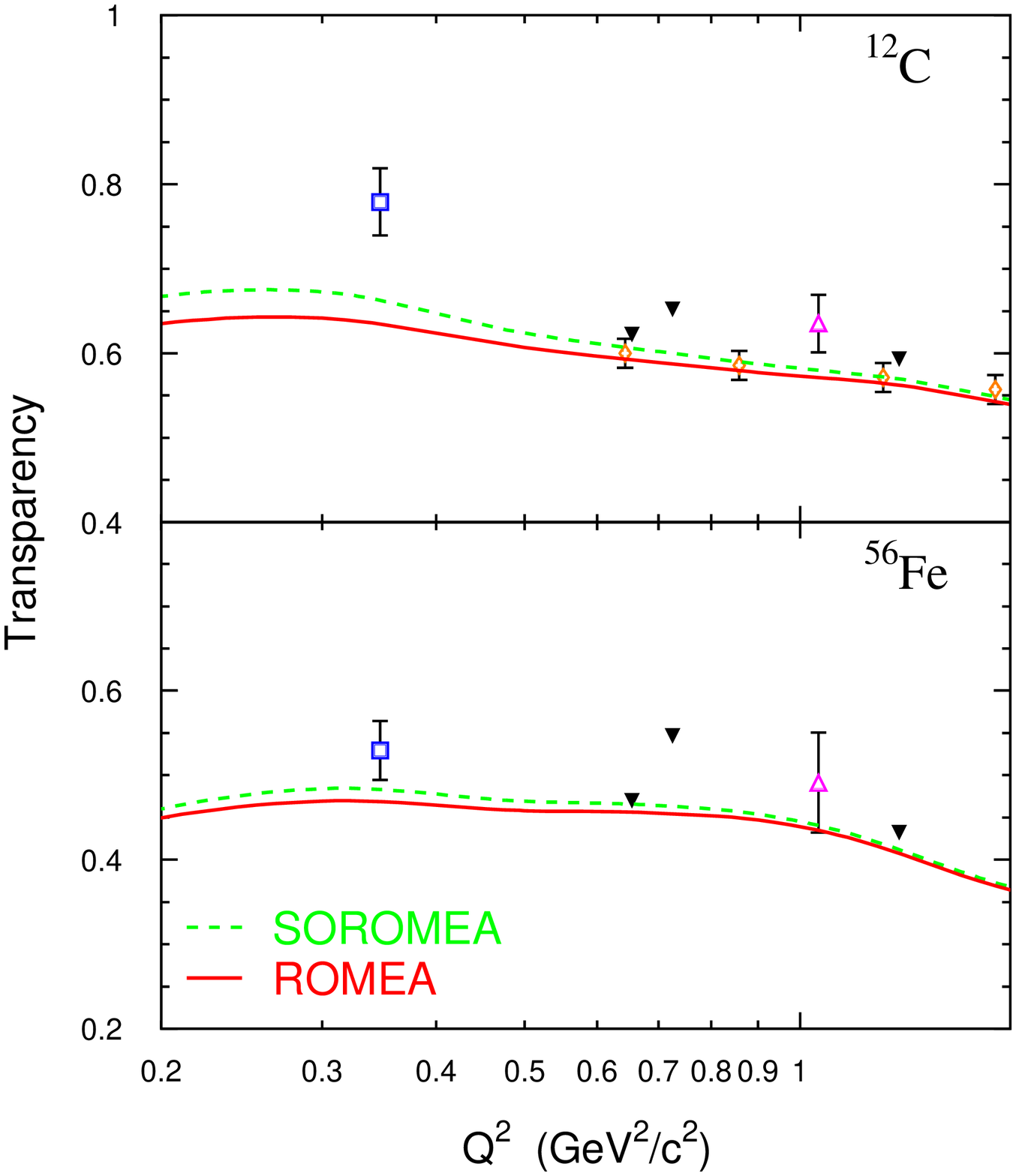}
\caption{Nuclear transparencies versus $Q^{2}$ for $A(e,e'p)$ reactions
in quasielastic kinematics.  The \mbox{SOROMEA} (dashed lines) are
compared to the \mbox{ROMEA} (solid lines) results.  The \mbox{EDAD1}
potential \cite{cooper93} has been employed in both formalisms.  Data
points are from Refs.~\cite{garino92} (open squares),
\cite{oneill95,makins94} (open triangles), \cite{abbott98,dutta03}
(solid triangles), and \cite{rohe05} (open diamonds).}
\label{fig:trans.eep.c12.fe56}
\end{center}
\end{figure*}

In Fig.~\ref{fig:trans.eep.c12.fe56} the \mbox{ROMEA} and \mbox{SOROMEA}
results are displayed as a function of the four-momentum transfer $Q^{2}$
and compared to the data.  Not surprisingly, at high $Q^{2}$, the
\mbox{ROMEA} and \mbox{SOROMEA} predictions practically coincide and the
role of the second-order eikonal effects grows with decreasing $Q^{2}$.
At $Q^{2} = 1.7$~(GeV/c)$^2$, the \mbox{ROMEA} and \mbox{SOROMEA}
transparencies agree to within $1\%$; while at $Q^{2} = 0.3$~(GeV/c)$^2$,
the difference has risen to $3\%$ for \nuc{56}{Fe} and $5\%$ for
\nuc{12}{C}.  The enhancement of the nuclear transparency due to the
second-order eikonal corrections is modest, even for values of the
four-momentum transfer as low as $Q^{2} = 0.2$~(GeV/c)$^2$.  Both the
\mbox{ROMEA} and the \mbox{SOROMEA} predictions tend to slightly
underestimate the measurements.  The second-order corrections move the
predictions somewhat closer to the $Q^{2} = 0.34$~(GeV/c)$^2$ data point.

As the nuclear transparency involves integrations over missing momenta
and energies, it may hide subtleties in the theoretical treatment of the
FSI mechanisms.  Next, we focus on highly exclusive $A(e,e'p)$ quantities
and quantify the role of second-order eikonal effects.

An observable that is particularly well suited to study \mbox{FSI} effects is
the induced normal polarization
\begin{equation}
P_{n} =
\frac{d^5 \sigma \left( \sigma_{n} = \uparrow \right) - 
d^5 \sigma \left( \sigma_{n} = \downarrow \right)}
{d^5 \sigma \left( \sigma_{n} = \uparrow \right) +
d^5 \sigma \left( \sigma_{n} = \downarrow \right)}
\; ,
\label{eq:eep_induced_pol}
\end{equation}
where $\sigma_{n}$ denotes the spin orientation of the ejectile in the
direction orthogonal to the reaction plane.  Indeed, in the one-photon
exchange approximation, $P_{n}$ vanishes in the absence of \mbox{FSI}.

\begin{figure*}[p]
\begin{center}
\includegraphics[width=0.75\textwidth]{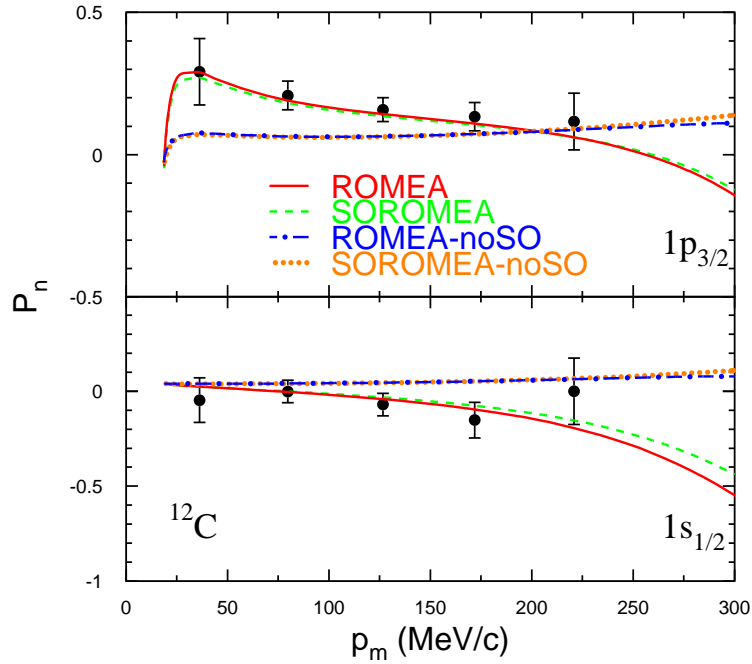}
\caption{Induced normal polarization $P_{n}$ for proton knockout from
the $1p_{3/2}$ (upper panel) and $1s_{1/2}$ (lower panel) shell in the
\nuc{12}{C}$(e,e'\vec{p})$ reaction.  The kinematics is determined by
beam energy $\epsilon = 579$~MeV, momentum transfer $q = 760$~MeV/c,
energy transfer $\omega = 292$~MeV, and azimuthal angle $\phi =
180^{\circ}$.  The solid (dashed) curves represent \mbox{ROMEA}
(\mbox{SOROMEA}) calculations.  The dot-dashed (dotted) curves refer
to predictions obtained within the \mbox{ROMEA} (\mbox{SOROMEA})
frameworks, with the spin-orbit term $V_{so} (\vec{b}, z) \,
\vec{\sigma} \cdot \vec{b} \times \vec{k}$ turned off.  The data are
from Ref.~\cite{woo98}.}
\label{fig:poln.12c.0.5.eep}
\end{center}
\end{figure*}

Fig.~\ref{fig:poln.12c.0.5.eep} shows the missing momentum dependence of
the induced normal polarization for the kinematics of Ref.~\cite{woo98},
corresponding with $Q^{2} \approx 0.5$~(GeV/c)$^2$.  The calculations are
performed with the energy-dependent $A$-independent (\mbox{EDAI})
potential of Ref.~\cite{cooper93}.  The \mbox{ROMEA} results are in line
with the relativistic distorted-wave impulse approximation (\mbox{RDWIA})
calculations of Ref.~\cite{udias00}.  The \mbox{RDWIA} framework was
implemented by the Madrid-Sevilla group \cite{udias93} and relies on a
partial-wave expansion of the exact scattering wave function.  It is
similar to the \mbox{(SO)ROMEA} approach in that both models compute the
effect of the \mbox{FSI} with the aid of proton-nucleus optical potentials.
Further, the overall agreement with the data is excellent.  The second-order
eikonal corrections are most pronounced for the $1s_{1/2}$ level.  For
missing momenta $p_m > 125$~MeV/c, they reduce the magnitude of the $P_{n}$
for the $1s_{1/2}$ state by roughly $20\%$, thereby resulting in a
marginally better agreement with the highest $p_m$ data point.  For
$1p_{3/2}$ knockout, on the other hand, the effect of the second-order
eikonal corrections is smaller than $5\%$.

The inclusion of the second-order eikonal effects is particularly visible
at high missing momentum, a region where also other mechanisms become
important.  The qualitative behavior of the meson-exchange and
$\Delta$-isobar currents, for instance, is alike \cite{ryckebusch99}.
At low missing momenta ($p_{m} \leq 200$~MeV/c), the induced normal
polarization $P_{n}$ is relatively insensitive to the two-body currents;
whereas at higher missing momenta, sizable contributions from the
meson-exchange and isobar currents are predicted.  The influence of the
meson and isobar degrees of freedom is also stronger for knockout from
the $1s_{1/2}$ shell than for $1p_{3/2}$ knockout.

In Fig.~\ref{fig:poln.12c.0.5.eep}, also calculations neglecting the
spin-orbit part $V_{so} (\vec{b}, z) \, \vec{\sigma} \cdot \vec{b} \times
\vec{k}$ are shown.  They illustrate that the spin-orbit distortion is
the largest source of $P_{n}$.  Hence, a correct inclusion of this term
is essential.  Moreover, $P_{n}$ proves to be rather sensitive to the choice
of optical potential \cite{udias00}.

Another $A(e,e'p)$ observable which has been the subject of many
investigations is the left-right asymmetry
\begin{equation}
A_{LT} = 
\frac{d^5 \sigma\left(\phi = 0^{\circ}\right) -
d^5 \sigma\left(\phi = 180^{\circ}\right)}
{d^5 \sigma\left(\phi = 0^{\circ}\right) +
d^5 \sigma\left(\phi = 180^{\circ}\right)} 
\; .
\label{eq:eep_A_LT}
\end{equation}
The subscript $LT$ indicates that this quantity is closely related to
the longitudinal-transverse response function.

\begin{figure*}[p]
\begin{center}
\includegraphics[width=0.75\textwidth]{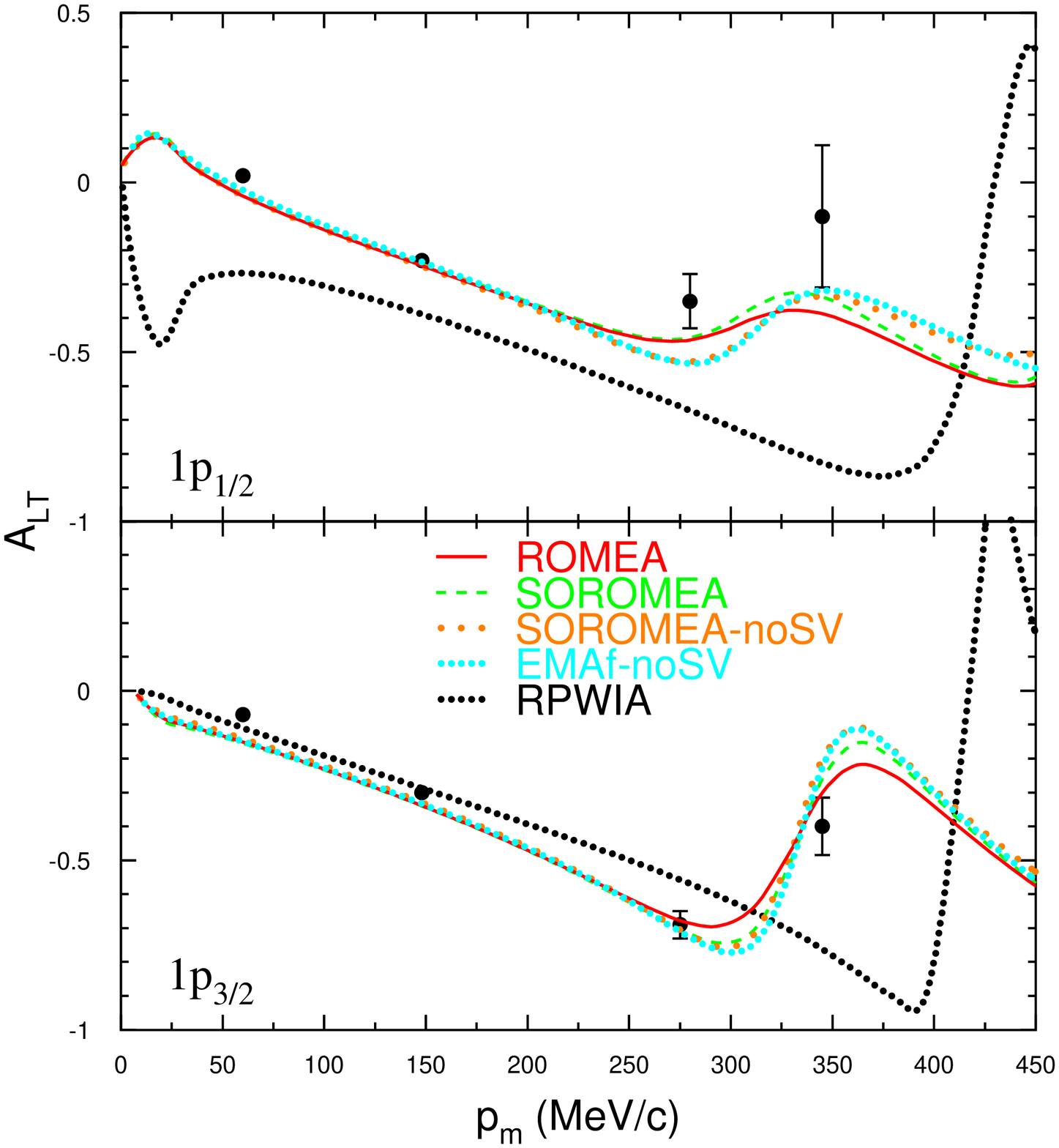}
\caption{The left-right asymmetry $A_{LT}$ for the \nuc{16}{O}$(e,e'p)$
experiment of \cite{gao00}.  The kinematics was $\epsilon = 2.442$~GeV,
$q = 1$~GeV/c, and $\omega = 445$~MeV (i.e., $Q^{2} = 0.8$~(GeV/c)$^{2}$).
The red solid (green dashed) lines show the results of the \mbox{ROMEA}
(\mbox{SOROMEA}) calculations.  The \mbox{SOROMEA-noSV} (orange
long-dotted curves) calculations differ from the \mbox{SOROMEA}
calculations in that the dynamical enhancement of the lower component
of the scattering wave function is neglected.  The cyan short-dotted
curves present the results from an \mbox{RDWIA} calculation where the
spinor distortions in the scattered wave are neglected.  All calculations
use the \mbox{EDAI} version for the optical potentials \cite{cooper93}.
The black short-dotted curves represent the \mbox{RPWIA} results.  The
data points are from Ref.~\cite{gao00}.}
\label{fig:atl.16o.0.8.eep}
\end{center}
\end{figure*}

Fig.~\ref{fig:atl.16o.0.8.eep} presents the $A_{LT}$ predictions for
the removal of $1p$-shell protons in \nuc{16}{O} in the kinematics of
Refs.~\cite{gao00,fissum04}.  The \mbox{FSI} shift the dip in $A_{LT}$,
which is located at $p_m \approx 400$~MeV/c in the relativistic
\mbox{PWIA} (\mbox{RPWIA}), to lower values of the missing momentum.
This shift is essential to describe the data at $p_m \approx 350$~MeV/c.
The exact $p_m$ location and height of the ripple, however, are affected
by many ingredients of the calculations, such as the current operator,
bound-state wave function, and parametrization of the optical potential
\cite{fissum04}.  As can be inferred from Fig.~\ref{fig:atl.16o.0.8.eep},
the second-order eikonal corrections affect the height, but not the
position of the ripple.

We also show the results of our \mbox{SOROMEA} calculations within the
so-called \mbox{noSV} approximation.  In this approximation, the dynamical
enhancement of the lower component of the scattering wave
(\ref{eq:lower_upper_relation}) due to the $V_{s} (r) - V_{v} (r)$ term
is omitted.  As such, the \mbox{SOROMEA-noSV} calculations make the same
set of assumptions as the \mbox{EMAf-noSV} predictions by the
Madrid-Sevilla group.  The \mbox{EMAf-noSV} approach is an \mbox{RDWIA}
calculation which adopts the EMA in combination with the \mbox{noSV}
approximation.  The second-order eikonal corrections clearly increase
the height of the oscillation in $A_{LT}$ and brings the eikonal
\mbox{noSV} calculations in excellent agreement with the corresponding
partial-wave prediction \mbox{EMAf-noSV}.  Finally, the comparison
between the \mbox{SOROMEA} and the \mbox{SOROMEA-noSV} calculations
demonstrates that the dynamical enhancement plays a significant role
in the description of the $A_{LT}$ data.

\begin{figure*}[p]
\begin{center}
\includegraphics[width=\linewidth]{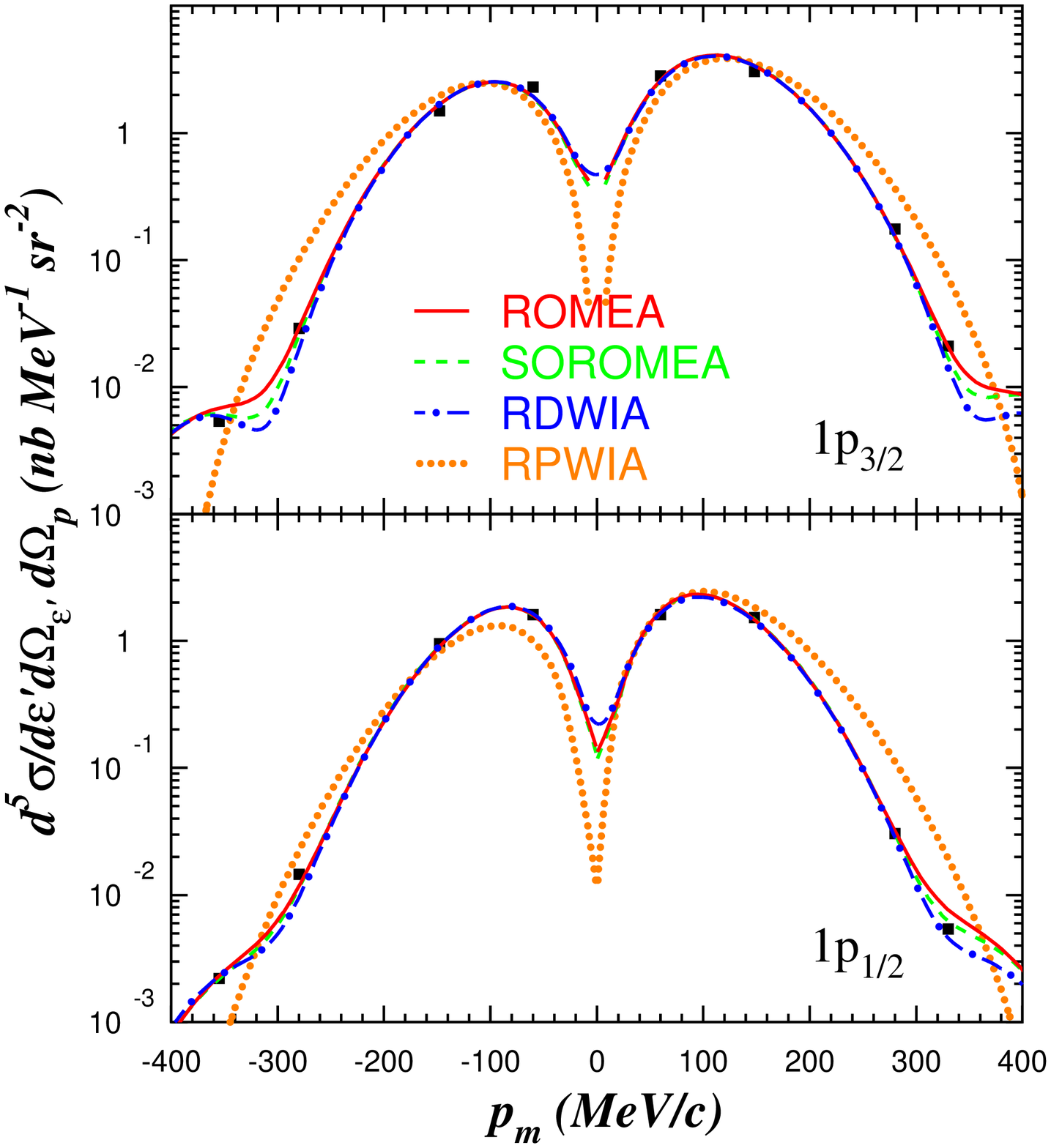}
\caption{\nuc{16}{O}$(e,e'p)$ cross sections compared to \mbox{ROMEA},
\mbox{SOROMEA}, \mbox{RDWIA}, and \mbox{RPWIA} calculations for the
constant $(\vec{q}, \omega)$ kinematics of Fig.~\ref{fig:atl.16o.0.8.eep}.
The calculations use the optical potential \mbox{EDAI} \cite{cooper93}.
The data are from Ref.~\cite{gao00} and the \mbox{RDWIA} results from
Ref.~\cite{fissum04}.  The following convention is adopted: positive
(negative) $p_m$ corresponds to $\phi = 180^{\circ}$ ($\phi = 0^{\circ}$).}
\label{fig:cross.16o.0.8.eep}
\end{center}
\end{figure*}

In Fig.~\ref{fig:cross.16o.0.8.eep}, \nuc{16}{O}$(e,e'p)$ cross-section
results are displayed for the kinematics of Fig.~\ref{fig:atl.16o.0.8.eep}.
The spectroscopic factors, which normalize the calculations to the data,
were determined by performing a $\chi^{2}$ fit to the data and are
summarized in Table~\ref{table:spec.16o.0.8.eep}.  The \mbox{RDWIA}
spectroscopic factors are $5$--$10\%$ higher than the \mbox{(SO)ROMEA}
ones.  The second-order eikonal corrections hardly affect the values of
the extracted spectroscopic factors.  Both our \mbox{(SO)ROMEA}
calculations and the \mbox{RDWIA} predictions of the Madrid-Sevilla group
do a very good job of representing the data over the entire $p_m$ range.
For missing momenta $|p_m| \leq 250$~MeV/c, the \mbox{(SO)ROMEA} and
\mbox{RDWIA} results are in excellent agreement.  The impact of the
second-order eikonal corrections on the computed differential cross
sections is almost negligible for $p_m$ below the Fermi momentum, but
can be as large as $30\%$ at high $p_m$.  The inclusion of the
second-order effects improves the agreement with the \mbox{RDWIA}
calculations at these high missing momenta.  Results for the effective
response functions $R_L$, $R_T$, $R_{LT}$, and $R_{TT}$ are not shown,
but the effect of the second-order eikonal corrections is similar to
the effect on the differential cross section.

\begin{table}[h]
\centering
\begin{tabular}{|c||c|c|c|c|}
\hline
  & \mbox{RPWIA}  & \mbox{ROMEA} & \mbox{SOROMEA} & \mbox{RDWIA} \\
\hline
\hline
$1p_{3/2}$ & 0.55 & 0.84 & 0.83 & 0.92 \\
\hline
$1p_{1/2}$ & 0.47 & 0.75 & 0.74 & 0.78 \\
\hline
\end{tabular}
\caption{The spectroscopic factors for the \nuc{16}{O}$(e,e'p)$ reaction
of Ref.~\cite{gao00}, as obtained with a $\chi^{2}$ procedure.}
\label{table:spec.16o.0.8.eep}
\end{table}

\section{Conclusions}
\label{sec:concl}

We have developed a formalism to account for second-order corrections in
the eikonal approximation.  Our model is relativistic and includes both
the central and spin-orbit parts of the optical potentials.  The formalism
has been applied to $A(e,e'p)$ processes.  Our numerical calculations
show that the effect of the second-order eikonal corrections on $A(e,e'p)$
observables is rather limited for $Q^{2} \geq 0.2$~(GeV/c)$^{2}$.  The
nuclear transparency calculations confirm the expected energy dependence
of the eikonal corrections: the effect decreases with increasing $Q^{2}$.
Concerning the $p_m$ dependence of the $A(e,e'p)$ observables, the effect
of the second-order eikonal corrections is minor except at high
missing momenta.  In this high-$p_m$ region, the eikonal corrections
affect the observables up to an order of $30\%$, thereby bringing the
calculations closer to the data and/or the \mbox{RDWIA} calculations.
The robustness of the first-order eikonal approximation, which emerges
from this study, can be invoked to explain the success of the Glauber
approach to $A(e,e'p)$ down to relatively low kinetic energies of
$200$~MeV.

\begin{ack}
This work was supported by the Fund for Scientific Research, Flanders (FWO).
\end{ack}




\begin{thebibliography}{00}
\bibitem{mccauley58}
G.P. McCauley, G.E. Brown, Proc. Phys. Soc. London 71 (1958) 893.
\bibitem{glauber59}
R.J. Glauber, in: W.E. Brittin, et al. (Eds.), Lectures in Theoretical
Physics, Interscience, New York, 1959.
\bibitem{joachain75}
C.J. Joachain, \textit{Quantum Collision Theory} (Elsevier, Amsterdam,
1975).
\bibitem{greenberg94}
W.R. Greenberg, G.A. Miller, Phys. Rev. C 49 (1994) 2747.
\bibitem{bianconi95}
A. Bianconi, M. Radici, Phys. Lett. B 363 (1995) 24.
\bibitem{ito97}
H. Ito, S.E. Koonin, R. Seki, Phys. Rev. C 56 (1997) 3231.
\bibitem{debruyne00}
D. Debruyne, J. Ryckebusch, W. Van Nespen, S. Janssen, Phys. Rev. C 62
(2000) 024611.
\bibitem{frankfurt95}
L.L. Frankfurt, E. Moniz, M. Sargsyan, M.I. Strikman, Phys. Rev. C 51
(1995) 3435.
\bibitem{nikolaev95}
N.N. Nikolaev, A. Szcurek, J. Speth, J. Wambach, B.G. Zakharov,
V.R. Zoller, Nucl. Phys. A 582 (1995) 665.
\bibitem{jeschonnek99}
S. Jeschonnek, T.W. Donnelly, Phys. Rev. C 59 (1999) 2676.
\bibitem{kohama00}
A. Kohama, K. Yazaki, R. Seki, Nucl. Phys. A 662 (2000) 175.
\bibitem{benhar00}
O. Benhar, N. Nikolaev, J. Speth, A. Usmani, B. Zakharov,
Nucl. Phys. A 673 (2000) 241.
\bibitem{ciofi00}
C. Ciofi degli Atti, L.P. Kaptari, D. Treleani, Phys. Rev. C 63 (2001)
044601.
\bibitem{Petraki2003}
M. Petraki, E. Mavrommatis, O. Benhar, J.W. Clark, A. Fabrocini,
S. Fantoni, Phys. Rev. C 67 (2003) 014605.
\bibitem{ryckebusch03}
J. Ryckebusch, D. Debruyne, P. Lava, S. Janssen, B. Van Overmeire,
T. Van Cauteren, Nucl. Phys. A 728 (2003) 226.
\bibitem{wallace}
S.J. Wallace, Phys. Rev. Lett. 27 (1971) 622;

S.J. Wallace, Ann. Phys. (N.Y.) 78 (1973) 190;

S.J. Wallace, Phys. Rev. D 8 (1973) 1846;

S.J. Wallace, J.A. McNeil, Phys. Rev. D 16 (1977) 3565;

S.J. Wallace, Phys. Rev. C 29 (1984) 956.
\bibitem{waxman81}
D. Waxman, C. Wilkin, J.-F. Germond, R.J. Lombard, Phys. Rev. C 24
(1981) 578.
\bibitem{faldt92}
G. F\"aldt, A. Ingemarsson, J. Mahalanabis, Phys. Rev. C 46 (1992)
1974.
\bibitem{carstoiu93}
F. Carstoiu, R.J. Lombard, Phys. Rev. C 48 (1993) 830.
\bibitem{cha95}
M.H. Cha, Y.J. Kim, Phys. Rev. C 51 (1995) 212.
\bibitem{alkhalili97}
J.S. Al-Khalili, J.A. Tostevin, J.M. Brooke, Phys. Rev. C 55 (1997)
R1018.
\bibitem{aguiar97}
C.E. Aguiar, F. Zardi, A. Vitturi, Phys. Rev. C 56 (1997) 1511.
\bibitem{tjon06}
J.A. Tjon, S.J. Wallace, Phys. Rev. C 74 (2006) 064602.
\bibitem{baker72}
A. Baker, Phys. Rev. D 6 (1972) 3462.
\bibitem{kelly96}
J.J. Kelly, Adv. Nucl. Phys. 23 (1996) 75.
\bibitem{serot86}
B.D. Serot, J.D. Walecka, Adv. Nucl. Phys. 16 (1986) 1.
\bibitem{furnstahl97}
R.J. Furnstahl, B.D. Serot, H.-B. Tang, Nucl. Phys. A 615 (1997) 441.
\bibitem{forest83}
T. de Forest, Nucl. Phys. A 392 (1983) 232.
\bibitem{amado83}
R.D. Amado, J. Piekarewicz, D.A. Sparrow, J.A. McNeil, Phys. Rev. C 28
(1983) 1663.
\bibitem{kelly99}
J.J. Kelly, Phys. Rev. C 60 (1999) 044609.
\bibitem{garino92}
G. Garino, et al., Phys. Rev. C 45 (1992) 780.
\bibitem{oneill95}
T.G. O'Neill, et al., Phys. Lett. B 351 (1995) 87.
\bibitem{makins94}
N.C.R. Makins, et al., Phys. Rev. Lett. 72 (1994) 1986.
\bibitem{abbott98}
D. Abbott, et al., Phys. Rev. Lett. 80 (1998) 5072.
\bibitem{dutta03}
D. Dutta, et al., Phys. Rev. C 68 (2003) 064603.
\bibitem{rohe05}
D. Rohe, et al., Phys. Rev. C 72 (2005) 054602.
\bibitem{lava04}
P. Lava, M.C. Mart\'{i}nez, J. Ryckebusch, J.A. Caballero,
J.M. Ud\'{i}as, Phys. Lett. B 595 (2004) 177.
\bibitem{cooper93}
E.D. Cooper, S. Hama, B.C. Clark, R.L. Mercer, Phys. Rev. C 47 (1993)
297.
\bibitem{woo98}
R.J. Woo, et al., Phys. Rev. Lett. 80 (1998) 456.
\bibitem{udias00}
J.M. Ud\'{i}as, J.R. Vignote, Phys. Rev. C 62 (2000) 034302.
\bibitem{udias93}
J.M. Ud\'{i}as, P. Sarriguren, E. Moya de Guerra, E. Garrido,
J. A. Caballero, Phys. Rev. C 48 (1993) 2731.
\bibitem{ryckebusch99}
J. Ryckebusch, D. Debruyne, W. Van Nespen, S. Janssen, Phys. Rev. C 60
(1999) 034604.
\bibitem{gao00}
J. Gao, et al., Phys. Rev. Lett. 84 (2000) 3265.
\bibitem{fissum04}
K.G. Fissum, et al., Phys. Rev. C 70 (2004) 034606.
\end{thebibliography}
\end{document}